\documentclass{cup-hpl}

\usepackage{lipsum}

\begin{document}

\newtheorem{theorem}{Theorem}

\shorttitle{Highly efficient difference-frequency generation}                                   
\shortauthor{K. Huang et al.}

\title{Highly efficient difference-frequency generation for mid-infrared pulses by passively synchronous seeding}

\author[1] {Kun Huang\corresp{K. Huang and H. Zeng, No. 500 Dongchuan Road, Shanghai 200241, China. \email{khuang@lps.ecnu.edu.cn (K. Huang); hpzeng@phy.ecnu.edu.cn (H. Zeng)}}}
\author[1]{Yinqi Wang}
\author[1]{Jianan Fang}
\author[2,3]{Huaixi Chen}
\author[4]{Minghang Xu}
\author[4]{Qiang Hao}
\author[1]{Ming Yan}
\author[1,5,6,7] {Heping Zeng}
\address[1]{State Key Laboratory of Precision Spectroscopy, East China Normal University, Shanghai 200062, China}
\address[2]{Key Laboratory of Optoelectronic Materials Chemistry and Physics, Fujian Institute of Research on the Structure of Matter, Chinese Academy of Sciences, Fuzhou, Fujian 350002, China}
\address[3]{University of Chinese Academy of Sciences, Beijing 100049, China}
\address[4]{Shanghai Key Laboratory of Modern Optical System, and Engineering Research Center of Optical Instrument and System, Ministry of Education, School of Optical Electrical and Computer Engineering, University of Shanghai for Science and Technology, Shanghai 200093, China}
\address[5]{Jinan Institute of Quantum Technology, Jinan, Shandong 250101, China}
\address[6]{CAS Center for Excellence in Ultra-intense Laser Science, Shanghai 201800, China}
\address[7]{Shanghai Research Center for Quantum Sciences, Shanghai 201315, China}

\begin{abstract}
We have proposed and experimentally demonstrated a novel scheme for efficient mid-infrared difference-frequency generation based on passively synchronized fiber lasers. The adoption of coincident seeding pulses in the nonlinear conversion process could substantially lower the pumping threshold for mid-infrared parametric emission. Consequently, a picosecond mid-infrared source at 3.1 $\mu$m was prepared with watt-level average power, and a maximum power conversion efficiency of 77\% was realized from pump to down-converted light. Additionally, the long-term stability of generated power was manifested with a relative fluctuation as low as 0.17\% over one hour. Thanks to the all-optical passive synchronization and all-polarization-maintaining fiber architecture, the implemented laser system was also featured with simplicity, compactness and robustness, which would favor subsequent applications beyond laboratory operation.
\end{abstract}


\maketitle

Mid-infrared (MIR) spectrum covers several transparent windows of the Earth's atmosphere, and accommodates so-called fingerprint region for molecular rotational-vibrational transitions. These unique features render MIR laser sources in great demand for a variety of scientific, industrial and medical applications including environmental sensing, atmospheric communication, molecular spectroscopy, material processing, microsurgical treatment, and biological analysis \cite{EbrahimZadeh2008Book,Schliesser2012NP,Serebryakov2010JOT}. So far, various approaches have been developed for direct MIR generation, such as supercontinuum generation \cite{Yu2013OME,Liu2014OE}, optical parametric oscillators (OPOs) \cite{Muraviev2018NP,Gu2013OL,Liu2019OL}, quantum cascaded lasers \cite{Yao2012NP}, and rare-earth doped fluoride fiber lasers \cite{Jackson2012NP,Zhu2017JOSAB,Ma2019APR}. 

Another common way to produce MIR light could resort to difference-frequency generation (DFG) between two near-infrared (NIR) beams. Notably, the DFG technique is featured with single-pass configuration, simple light-path alignment and reduced number of optical components, which thus provides desirable advantages like compactness, robustness and versatility \cite{Dunn1999Science}. Additionally, it has been shown that simultaneous injection of signal and pump fields could significantly lower the required pump power to approach high-power and/or high-efficiency performance \cite{Xu2015OE,Belden2015OL,Yue2020OL}. Consequently, the reduced pumping threshold for MIR parametric emission would alleviate the damage risk of nonlinear crystal. Over past decades, the nonlinear parametric effect has been exploited to prepare MIR coherent light of superior optical quality with temporal coverage ranging from continuous wave to femtosecond pulses \cite{Dunn1999Science}.

\begin{figure*}[t!]
\centering
\includegraphics[width=0.8 \textwidth]{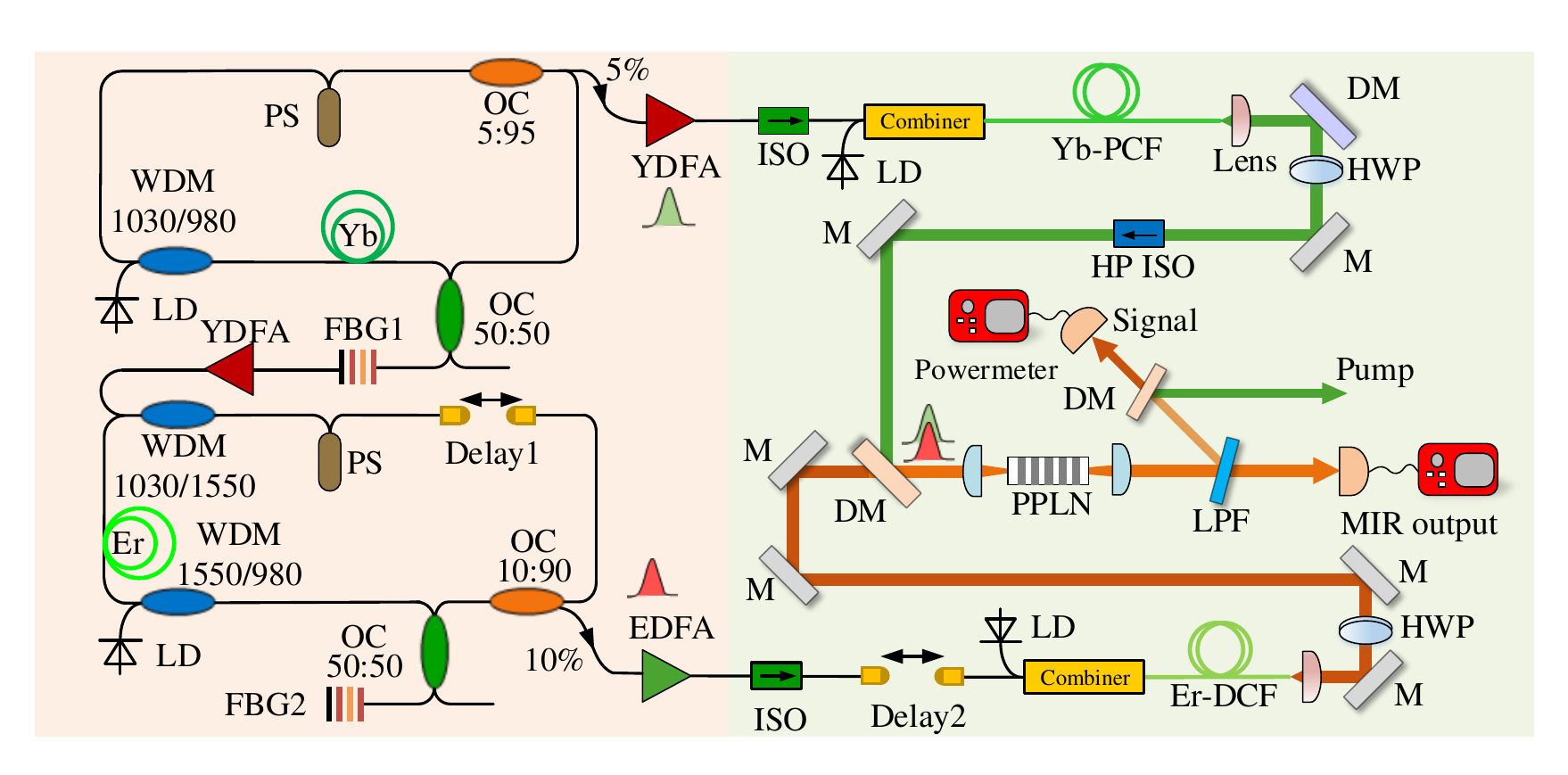}
\caption{Experimental schematic for mid-infrared generation based on passively synchronized ultrafast fiber laser system. The pump and signal pulses originated from mode-locked Yb- and Er-doped fiber lasers, respectively. After cascaded fiber amplifiers, the two-color pulses were steered into a PPLN crystal for implementing difference-frequency generation. Consequently, the average power and conversion efficiency for the MIR output could be effectively improved due to the synchronous seeding. LD: laser diode; WDM: wavelength division multiplexer; Yb/Er: ytterbium/erbium-doped gain fiber; OC: optical coupler; PS: phase shifter; FBG: fiber Bragg grating; Col: collimator; PCF:  Photonic crystal fiber; DCF: double-clad fiber; DM: dichroic mirror; HWP: half-wave plate; M: mirror; HP ISO: high-power isolator; LPF: long-pass filter; PPLN: periodically-poled lithium niobate crystal.}
\label{fig1}
\end{figure*}

In particular, the DFG technique for MIR pulsed generation has been greatly fueled by the maturation of efficient periodically poled nonlinear crystals, as well as the advances on high-power fiber lasers and amplifiers in the ultrafast regime. To optimize the nonlinear conversion efficiency, it is prerequisite to temporally synchronize the two NIR pulses with a sufficiently small relative timing jitter. Although the synchronized dual-color light sources could be accessed based on supercontinuum generation \cite{Hu2017AO, Oliveira2020OL} or Raman-induced soliton self-frequency shift \cite{Sobon2017OL} from the same laser source, yet the available power at specified wavelength windows was usually limited by the conversion efficiency. Moreover, the nonlinear spectral extension between two disparate wavelengths might inevitably suffer from detrimental intensity instability or degraded optical coherence due to the competition of various nonlinear processes \cite{Steinle2014OE}. 

Alternatively, two-color synchronized pulses could be obtained from two independent laser sources by tightly locking their relative repetition rates \cite{Tian2017IEEE}. For instance, two independent laser sources have been used to achieve watt-level high-efficiency mid-infrared generation by using two amplitude modulators with a common timing clock, albeit that the available output pulse duration was limited by the bandwidth of electronic devices \cite{Murray2016OL}. To generate much shorter MIR pulses, active synchronization system based on two ultrafast mode-locked lasers was exploited \cite{Xuan2012APB}. Recently, there has been emerging investigation on all-optical passive synchronization between ultrafast fiber lasers based on cross-phase modulation \cite{Huang2018OE,Li2020OL}. The passive fashion of timing locking could effectively mitigate the system complexity in the active configuration, which may facilitate synchronously pumping DFG.

In this work, we proposed and implemented a novel scheme for ultrafast MIR generation based on an all-optical passive synchronization fiber laser system, which eliminated the stringent requirement of complicated feed-back system and high-speed electronics. In combination with techniques of spectro-temporal pulse engineering and high-power fiber amplifiers, we finally obtained 1.24-W MIR output with a high conversion efficiency about 77\%. Additionally, thanks to the all-polarization-maintaining fiber structure, the whole system was also featured with compact layout and long-term stability.

Figure \ref{fig1} illustrates the schematic diagram of experimental setup for synchronously pumping MIR generation, which was mainly comprised of two parts, \textit{i.e.,} synchronized fiber lasers and difference-frequency generation. The synchronized fiber laser system included two Yb- and Er-doped fiber lasers (YDFL and EDFL) in a master-slave layout \cite{Huang2018OE}. The configuration of the fiber lasers were based on nonlinear amplifying loop mirrors (NALMs). The counter-circulating pulses within the loop would interfere on the the symmetric optical coupler. The transmission of the Sagnac-type interferometer depended on the relative phase between the bidirectional pulses. The NALM thus acted as an effective saturable mirror, which would initiate the passive mode-locking for ultrashort pulse generation at a repetition rate about 20.1 MHz. The phase-shifter (PS) was used to provide a $\pi/2$ linear non-reciprocal phase within the loop, which could substantially lower the pump threshold for the mode-locking operation. The fiber Bragg grating (FBG) was used as the end mirror for the laser resonator. The bandwidth of the FBG enabled us to control the output spectrum. The delay line (Delay1) in the EDFL was constructed with a pair of fiber collimators, where one was mounted on a translational stage with precision of 50 $\mu$m.

In order to obtain the synchronization between the two-color fiber lasers, the transmitted portion of the FBG in the master laser was injected into the slave laser cavity after being amplified via a Yb-doped fiber amplifier (YDFA). The passive locking of the relative repetition rate was realized by the cross-phase modulation effect between the master injection and slave pulses. The underlying mechanism for the all-optical synchronization lied in the effective fast intensity modulation due to the periodic introduction of nonreciprocal phase difference within the phase-biased Sagnac interferometer loop \cite{Huang2018OE}. In contrast to previous schemes, the synchronization system was configured in an all-polarization maintaining (all-PM) structure, thus gaining substantially improved stability and robustness. Notably, the tolerance range of the cavity-length mismatch could reach to the centimeter level, which was essential to maintain the long-term stable performance.

\begin{figure}[b!]
\centering
\includegraphics[width=1\columnwidth]{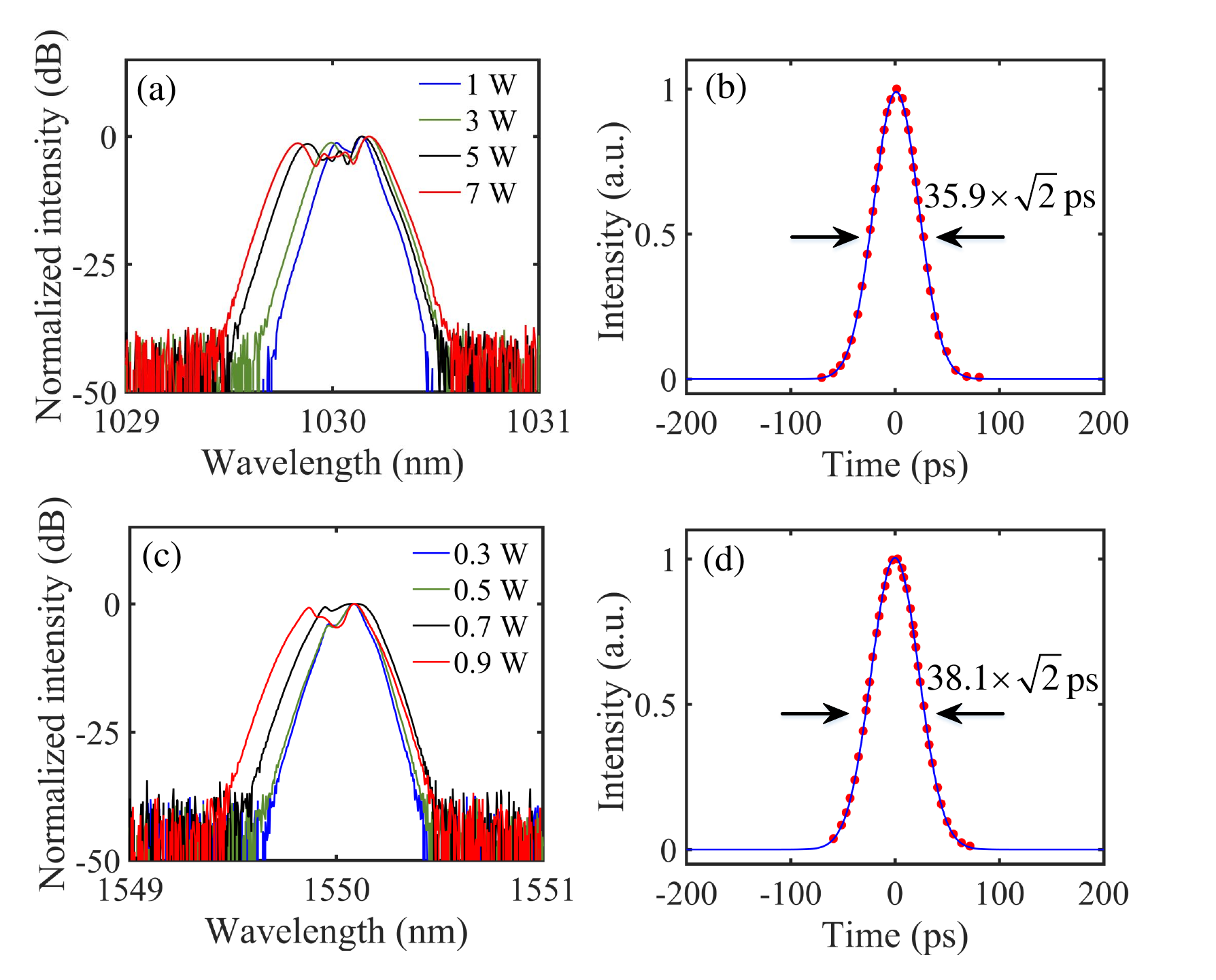}
\caption{ Experimental characterization of output pulses from pump (a, b) and signal (c, d) after two-stage fiber amplifiers, including the measured optical spectra (a, c) and corresponding auto-correlation traces (b, d). Note that the actual intensity profiles are scaled down by a factor of $\sqrt{2}$ under an assumption of Gaussian pulses.}
\label{fig2}
\end{figure}

Then the synchronized dual-color pulses from YDFL and EDFL were steered into amplification stages for power boosting, which would serve as pump and signal sources in the subsequent DFG. The pump average power could reach to 7 W after a single-mode preamplifier and main amplifier based on photonic crystal fiber (PCF, NKT Photonics DC-135/14-PM-YB). The resulting spectrum was shown in Fig. \ref{fig2}(a), indicating a center wavelength at 1030 nm and a 3-dB bandwidth of 0.47 nm. The broadening and oscillating behavior as increasing the output power was ascribed to non-negligible self-phase modulation within the fiber waveguide. The corresponding auto-correlation trace was given in Fig. \ref{fig2}(b), which implied a pulse duration about 35.9 ps by assuming a Gaussian profile. In parallel, the signal pulse was boosted by a double-clad fiber (DCF, Nufern PM-EYDF-12/130-HE) amplifier to a maximum average power of 900 mW. The spectrum was centered at 1550 nm with a 3-dB bandwidth of 0.34 nm. The pulse duration was inferred to be 38.1 ps. The comparable pulse durations between signal and pump sources were chosen to optimize the nonlinear conversion efficiency \cite{Murray2016OL}. It is also worth to noting that the relative timing jitter between the pump and signal pulses was typically about tens of femtoseconds \cite{Huang2018OE}. The achieved tight synchronization was essential to ensure efficient and stable nonlinear interaction.

\begin{figure}[t!]
\centering
\includegraphics[width=1\columnwidth]{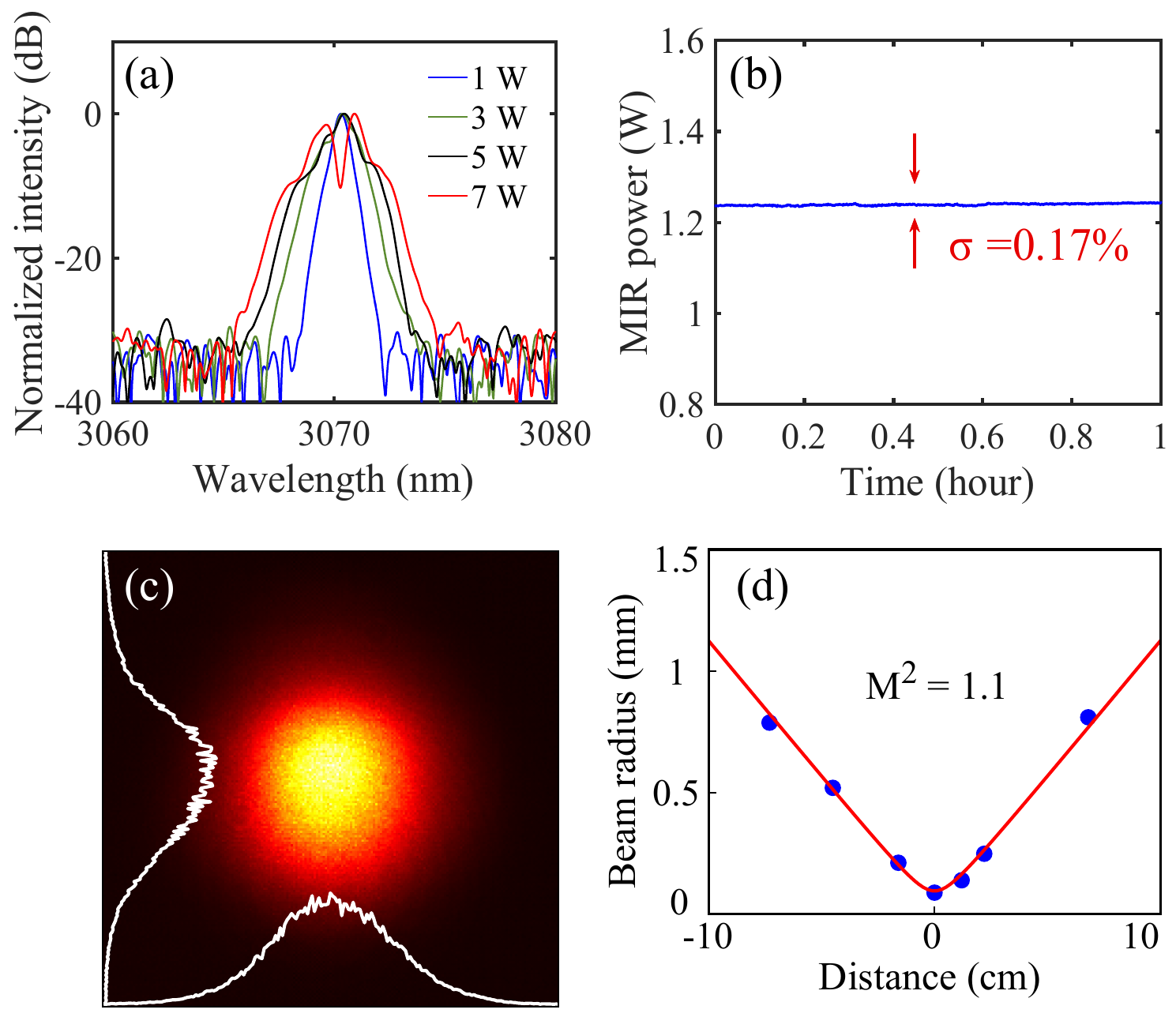}
\caption{(a) MIR spectra under different pump power. The signal power was kept at 900 mW. (b) Power stability of the mid-infrared output. $\sigma$ indicates the relative fluctuation. (c) MIR beam image at the near field as well as two section profiles along orthogonal axes. (d) Evolution of MIR beam waists along the propagation distance. Note that the central position was defined at the focal point. }
\label{fig3}
\end{figure}

Finally, the two-color pulses were spatially combined by a dichroic mirror (DM1) and temporally overlapped by a delay line (Delay2). Their polarizations were adjusted to be vertical by half-wave plates to satisfy the type-0 phase-matching condition. The combined beam was then focused by achromatic lens into a periodically-poled lithium niobate (PPLN) crystal with a length of 25 mm and a thickness of 1 mm. Both end surfaces of the nonlinear crystal were antireflection coated for three relevant bands, \textit{i.e.}, 1030-1080 nm, 1380-1800 nm and 2400-4500 nm. The PPLN crystal was installed in an oven with a copper heat sink for implementing active temperature stabilization. The operation temperature was set at  37.8 $^\circ$C with a precision of 0.1 $^\circ$C, corresponding to a periodically-poling period of 30.3 $\mu$m used in our experiment. As shown in Fig. \ref{fig1}(a), the generated MIR light was collimated by a calcium fluoride (CaF$_2$) plano-convex lens with a focus length of 75 mm. The collimated MIR beam was then spectrally purified by another dichroic mirror (DM2) and a long-pass filter (LPF) with a cutoff wavelength of 2.4 $\mu$m. The power transmissions of the CaF$_2$ lens, DM2 and LPF for the MIR light at 3070 nm were about 90\%, 78\% and 89\%, respectively. Figure \ref{fig3}(a) presents the acquired MIR spectrum by a Fourier transform optical spectrum analyzers (Thorlabs OSA207C) with a resolution about 0.2 nm. The narrow-band spectrum enabled us to access a high-brightness MIR laser source, which would be desirable for applications such as material processing and free-space communication. Furthermore, the average power was recorded by a thermal power sensor (Thorlabs S401C), exhibiting a relative fluctuation as low as 0.17\% over one hour as shown in Fig. \ref{fig3}(b). The superior long-term stability was benefitted from the fact that all the fiber oscillators and amplifiers before nonlinear conversion were constructed with fully polarization-maintaining components. In combination with the passive all-optical synchronization, the whole laser system was in favor of compact layout and stable operation. Additionally, we also characterized the beam profile of the DFG light by using an infrared beam imager (Dataray,  WinCamD-IR-BB) with a high resolution of 17 $\mu$m. Figure \ref{fig3}(c) shows the acquired two-dimensional shape and cutting lines along two orthogonal directions, indicating a good beam quality. The spatial evolution was also measured as shown in Fig. \ref{fig3}(d), which resulted in M$^2$ about 1.1 in both directions. 

\begin{figure}[b!]
\centering
\includegraphics[width=0.8\columnwidth]{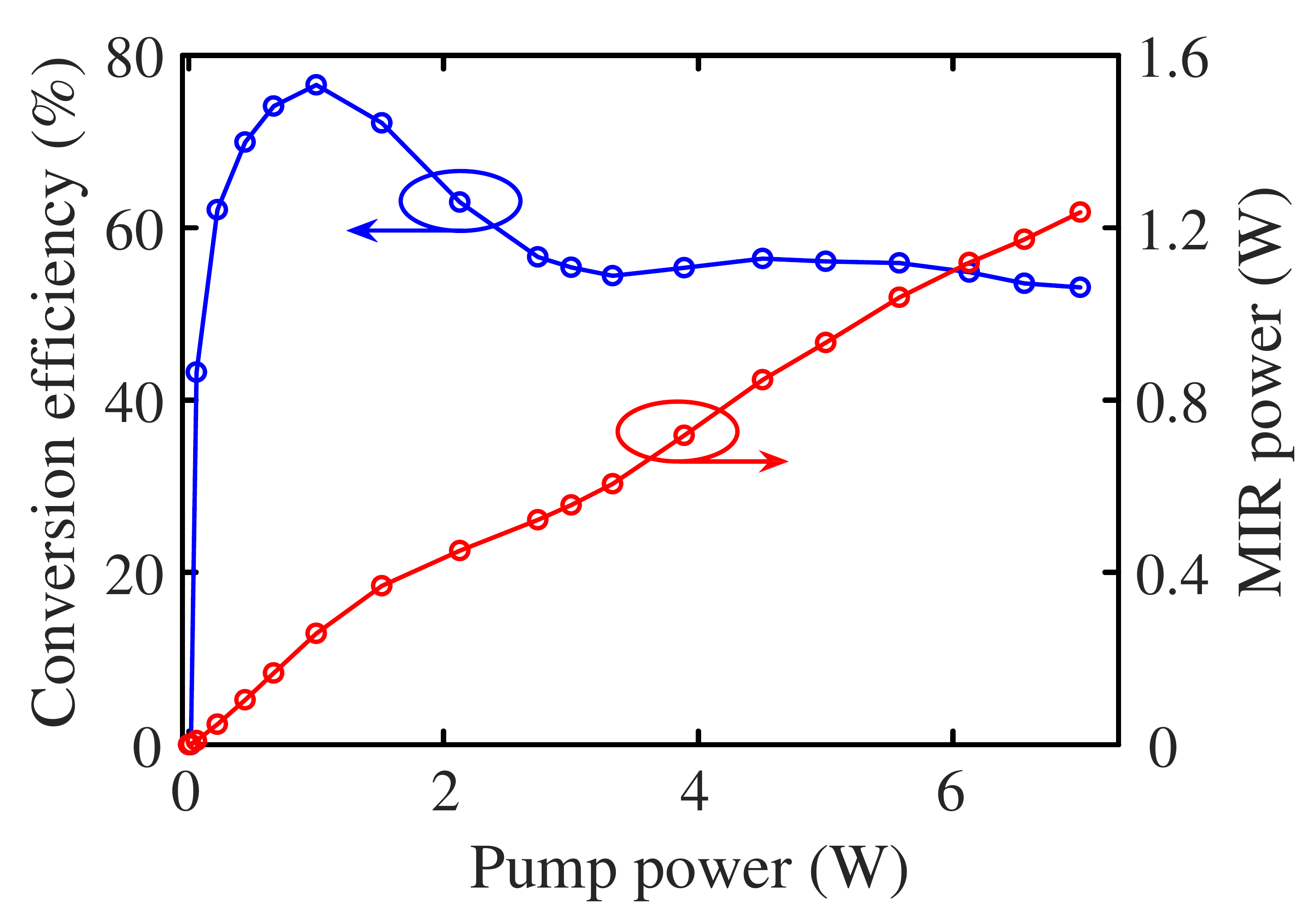}
\caption{Generated mid-infrared power and corresponding conversion efficiency vary as a function of the pump power. Note that the conversion efficiency was defined as total power of down-converted fields divided by the initial pump power. Connecting lines are only used to guide the eye.}
\label{fig4}
\end{figure}

Next we turn to investigate the DFG behavior as varying the power of involving optical fields. At a fixed signal power of 0.9 W, the MIR power increased monotonically with the increase of pump power, leading to maximum of 1.24 W. Note that the power value has been corrected by the total propagation transmission of 62\% ($90\% \times 78\% \times 89\%$) including the lens and filtering stage. We define the conversion efficiency $\eta$ as the ratio between the generated DFG total power (MIR power $P_\text{MIR}$ and signal power $P_\text{signal}$) and the pump power ($P_\text{pump}$), which could be expressed as
\begin{equation}
\eta  =  \frac{P_\text{MIR} + P_\text{signal}}{P_\text{pump}} = \frac{P_\text{MIR}}{P_\text{pump}} \times \frac{\lambda_\text{MIR}}{\lambda_\text{pump}}   \ ,
\label{eq1}
\end{equation}
where we have used the Manley-Rowe relation $P_\text{signal}/P_\text{MIR} = \lambda_\text{MIR}/\lambda_\text{signal}$, and the energy conversation law $1/\lambda_\text{pump} = 1/\lambda_\text{signal} + 1/\lambda_\text{MIR}$. As shown in Fig. \ref{fig4}, the conversion efficiency reached to a maximum of 77\%, and then started to decrease beyond a pump power about 1 W. The observed roll-off of conversion efficiency was attributed to the back-conversion of the signal and MIR power to the pump beam, which has also been reported in \cite{Murray2016OL}. Compared to the previous demonstration based on actively synchronized lasers \cite{Xuan2012APB}, both the conversion efficiency and output power here were much higher owing to  spectro-temporal matching and high-fidelity amplification in the experiment. It is noteworthy that without injecting the signal at 1.55 $\mu$m we could barely observe the MIR emission even at 7-W pumping, which indicated a typically high threshold to realize pronounced spontaneous down-conversion. Indeed, the synchronous seed injection could effectively induced the parametric generation \cite{Xu2015OE,Belden2015OL}. Therefore, the requirement of intense pump for high-power and/or high-efficiency MIR generation could be significantly relaxed, which would lower the damage risk for nonlinear crystals and mitigate the performance degradation due to thermal and photo-refractive effects. 

\begin{figure}[t!]
\centering
\includegraphics[width=0.75\columnwidth]{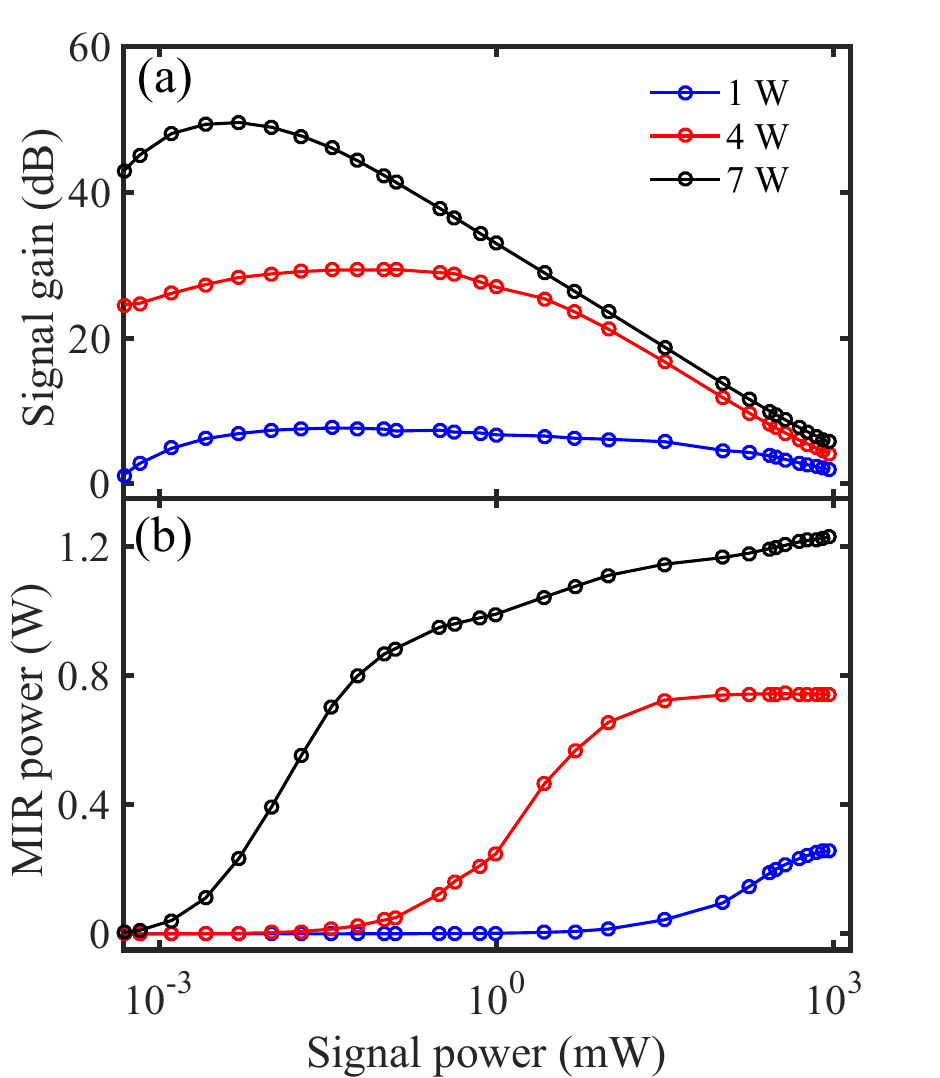}
\caption{(a) Generated MIR power versus injecting signal power under various settings of pump power. (b) Signal gain against input signal power for difference pump power. Note that connecting lines are used to guide the eye only.}
\label{fig5}
\end{figure}

Furthermore, we have characterized the DFG by varying the signal intensity for various pump settings. Figure \ref{fig5}(a) presents the influence of signal power on the parametric gain in the DFG process. In the case of a small signal power of 5 $\mu$W, the corresponding gain of the amplified signal could reach to 50 dB for a 7-W pump. In comparison to the previous demonstration \cite{Murray2016OL}, higher maximum signal gain could be obtained here at a much lower input power, which indicated the efficacy of the employed ultrashort pulse seeding. As illustrated in Fig. \ref{fig5}(b), the MIR power would tend to be saturated by augmenting the signal power, and the turning point for the saturation would be smaller under more intensive pumping. Specifically, at the presence of 7-W pump, watt-level MIR power could be produced with only mW-level signal injection.

To conclude, we have proposed and implemented a novel scheme for generating MIR ultrafast pulses, which relied on DFG between temporally synchronized dual-color mode-locked fiber lasers. Thanks to the passively synchronous seeding, the pump threshold for efficient parametric conversion was substantially reduced, which enabled us to obtain a maximum MIR output power of 1.24 W and a peak nonlinear conversion efficiency up to 77\%. Notably, the whole system benefited from the all-optical passive synchronization and all-polarization-maintaining configuration, which favors compact layout, self-starting operation and superior long-term stability. 

Therefore, the demonstrated system here would provide practical picosecond MIR light source for potential applications such as laser selective cutting of biological tissues \cite{Franjic2009OE} and plasma-free water droplet shattering for efficient fog clearing \cite{Rudenko2020Optica}. Moreover, the presented configuration might be engineered to access MIR pulses from femtosecond to nanosecond with the help of proper fiber-laser design and intra-cavity dispersion management. Also, direct combinations with wavelength-tuning or spectral-broadening techniques would permit to obtain MIR pulsed sources with a broadband spectral tunability, which would be promising for expanding fields.

\section*{Acknowledgments}
This work was supported in part by by National Key Research and Development Program (2018YFB0407100), Science and Technology Innovation Program of Basic Science Foundation of Shanghai (18JC1412000), Program for Professor of Special Appointment (Eastern Scholar) at Shanghai Institutions of Higher Learning, National Natural Science Foundation of China (11621404, 11727812), Shanghai Municipal Science and Technology Major Project (2019SHZDZX01).


\end{document}